\documentclass[10pt,conference]{IEEEtran}
\usepackage{dcolumn}
\usepackage{fancyheadings}
\usepackage{amstext,amssymb}
\usepackage{graphicx}
\usepackage{times,color}
\usepackage[final,ulem=normalem]{changes}
\usepackage{url}
\usepackage{subfigure}
\usepackage{algorithm}
\usepackage{algorithmic}
\usepackage{latexsym}
\usepackage{amsmath}
\usepackage{flushend}
\usepackage{epstopdf}

\newcommand{\comment}[1]{}

\definecolor{dgreen}{rgb}{0,0.48,0.3}

\long\def\comment#1{}

\IEEEoverridecommandlockouts

\begin{document}
\pagestyle{empty}
\title{Enhanced Precision Through Multiple Reads\\ for LDPC Decoding in Flash Memories
\thanks{Manuscript received May 15, 2013; revised October 3, 2013 and November 21, 2013.  This work was presented in part at Globecom 2011 in Houston, Texas in December 2011 and at the 2012 Non-Volatile Memories Workshop at UCSD in March of 2012.  This research was supported by a gift from Inphi Corp. and UC Discovery Grant 192837.  J. Wang is with Qualcomm Inc., San Diego, CA 92121 USA (email wjd1226@gmail.com).  K. Vakilinia and R. D.  Wesel are with The UCLA Electrical Engineering Department, Los Angeles, CA 90095 USA (emails: vakiliniak@ucla.edu, wesel@ee.ucla.edu). T.-Y. Chen is with the Northwestern University Electrical Engineering and Computer Science Department, Evanston, IL, 60208 USA (email tsungyi.chen@northwestern.edu).  T. Courtade is with the Department of Electrical Engineering and Computer Sciences, University of California, Berkeley, CA 94720 (email: courtade@eecs.berkeley.edu). G. Dong is with Skyera, Inc., San Jose, CA, 95131 USA (email dongguiqiang@gmail.com). Tong Zhang is with the Electrical, Computer, and Systems Engineering Department, Rensselaer Polytchnic Institute, Troy, NY 12180 USA (email tzhange@ecse.rpi.edu). Hari Shankar is with Inphi Corp. Westlake Village, CA 91362 USA (email hshankar@inphi.com). }}


\author{Jiadong Wang, Kasra Vakilinia, Tsung-Yi Chen, Thomas Courtade,  \\Guiqiang Dong, Tong Zhang,  Hari Shankar, and Richard Wesel}

\maketitle \thispagestyle{empty}

\begin{abstract}
Multiple reads of the same Flash memory cell with distinct word-line voltages provide enhanced precision for LDPC decoding.  In this paper, the word-line voltages are optimized by maximizing the mutual information (MI) of the quantized channel.  The enhanced precision from a few additional reads allows frame error rate (FER) performance to approach that of full-precision soft information and enables an LDPC code to significantly outperform a BCH code.  

A constant-ratio constraint provides a significant simplification in the optimization with no noticeable loss in performance.   

For a well-designed LDPC code, the quantization that maximizes the mutual information also minimizes the FER in our simulations.  However,  for an example LDPC code with a high error floor caused by small absorbing sets, the MMI quantization does not provide the lowest frame error rate. The best quantization in this case introduces more erasures than would be optimal for the channel MI in order to mitigate the absorbing sets of the poorly designed code.  

The paper also identifies a trade-off in LDPC code design when decoding is performed with multiple precision levels; the best code at one level of precision will typically not be the best code at a different level of precision.  

\end{abstract}



\section{Introduction}
Flash memory can store large quantities of data in a small device that has low power consumption and no moving parts. The original NAND Flash uses only two levels. This is called single-level-cell (SLC) Flash because there is only one actively written charge level. Devices currently available using four levels are called multi-level-cell (MLC) Flash. Four and eight levels are currently in use, and the number of levels will increase further \cite{LiISSCC08,TrinhISSCC08}.

Error control coding for Flash memory is becoming more important as the storage density increases. The increasing number of levels increases sensitivity to variations in signal-to-noise ratio (SNR) from cell to cell and over time due to wear-out.  This makes stronger error-correction codes necessary. Reductions in feature size make inter-cell interference more likely, adding an equalization or interference suppression component to the read channel \cite{LeeElectron02}. Also, the wear-out effect is time\added{-}varying, introducing a need for adaptive coding or modulation to maximize the potential of the system.

\subsection {Related Work}
Low-density parity-check (LDPC) codes are well-known for their capacity-approaching ability for AWGN channels \cite{RichardsonDes} and are the subject of recent interest for application to the Flash memory read channel.  For example, in \cite{YaakobiICCNC2012} LDPC codes without access to enhanced precision are shown to provide a performance improvement over BCH codes, but that improvement becomes small at high code rates.  Also in \cite{YaakobiICCNC2012}, an alternative error correction scheme is introduced that takes into account the dominant cell-level errors found in eight-level cells.   This scheme provides  improvement for eight-level cells without using enhanced precision.  

Important work related to codes that consider the dominant cell-level error is that of Gabrys et al. on graded bit error correcting codes \cite{GabrysIT2013}.  In contrast to codes designed for dominant errors, our paper focuses on the use of enhanced precision to improve performance.  While we explore the improvement in terms of standard LDPC codes, enhanced precision should also improve the performance of alternative error correction schemes that focus on the dominant cell-level errors as long as the decoders can utilize soft information.

Another approach for using LDPC codes in Flash memories \cite{ZhangISIT2010} is to design the codes for use with rank modulation.   Rank modulation \cite{JiangISIT2008,MazumdarISIT2011,QinISIT2013}  stores information in the cell using the {\em relative} value (or ordering) of cell charge levels rather than the absolute value.   LDPC codes for rank modulation require the cell charge-level ordering at the decoder.  

As observed in \cite{JiangISIT2008}, rank modulation eliminates the need for discrete cell levels, overcomes overshoot errors when programming cells, and mitigates the problem of asymmetric errors.  This is an exciting approach for future Flash architectures.  However, current Flash systems use the same word-line voltage to read all cells on the page and thus would require a large number of page reads to learn the charge-level ordering.  Our paper focuses on the traditional approach of coding with fixed target charge levels and assumes that when reading each page, the same word-line voltage is used for all cells.   

We note that an alternative to using multiple reads to enhance precision is to perform a single read with a dynamic threshold as introduced by \cite{SalaTCOM2013} to adapt to time-varying channel degradations such as the mean shift that occurs due to retention loss.  We note that dynamic thresholds are complementary to the use of enhanced precision, and a combined approach could be especially effective.

As in the precursor conference paper \cite{JWANGGLOBECOM11} this paper uses mutual information maximization as the objective function that drives the optimization of the word-line voltages (thresholds) used for the multiple reads that provide enhanced precision.  Mutual information maximization is also explored in \cite{LeeISIT05} for the design of memory-efficient decoding of LDPC codes and in \cite{KurkoskiIT} for quantization of binary-input discrete memoryless channels and the design of the message-passing decoders of LDPC codes used on such channels.

Another aspect of current research follows from the fact that Flash memory systems must erase an entire block of data at once.  Each block consists of numerous pages and each page contains thousands of bits.  Even to change a small amount of data on a single page, the entire block must be erased.  Moreover, the process of erasing and re-writing a block of data degrades performance.  Each time electrons are written and then erased from the floating gate, the integrity of the floating gate degrades in a process known as ``cell wear-out''.  

In \cite{BruckISIT2009}, coding is used to minimize the frequency with which a block must be erased and the number of auxiliary blocks required for moving pages of data in a Flash memory system.  Efficient wear-leveling and data movement in Flash is an important problem, but our paper addresses the complementary  problem of improving the ability to reliably read a page by using enhanced precision.

\subsection{Overview and Contributions}

LDPC codes have typically been decoded with soft information (a relatively high-precision representation of a real or complex number describing a received symbol value) while Flash memory systems have typically provided only hard reliability information (a single bit representing the output of a sense-amp comparator) to their decoders. This paper demonstrates that enhanced precision through multiple reads is crucial to successfully reaping the benefits of LDPC coding in Flash memory. We explore how to select the word-line voltages used for additional reads, how many such reads are necessary to provide most of the \replaced{LDPC performance benefit}{benefits}, and how varying levels of precision can impact code design.  

Section~\ref{background} briefly introduces the NAND Flash memory read channel model. Section~\ref{casestudy} shows how to obtain word-line voltages by maximizing the mutual information (MI) of the equivalent read channel using a simple Gaussian model of SLC (two-level) Flash as an example.  This section also shows that a few additional reads provide most of the benefit of enhanced precision through both a mutual information analysis and an LDPC simulation example.

Section \ref{sec:LDPCDesign} describes the LDPC codes used in the paper in detail.  This section also demonstrates a code design trade-off as follows: the best code in terms of both density evolution threshold \cite{RichardsonDes} and empirical performance at one precision level is not the best according to either density evolution threshold or empirical performance at another precision level.  This is a practically important issue because the same code may well be decoded with varying levels of precision.  In a practical system it is likely that additional page reads to enhance precision will be used only if the page could not be decoded without them.   

Section \ref{sec:4-levels} extends the discussion to MLC (four-level) Flash.   This section uses a more realistic model of the Flash read channel from \cite{DongUSENIX2011} and employs the ``constant-ratio'' method of \cite{DongTCAS2011} as a constraint to simplify the threshold optimization.  This section confirms that maximizing mutual information also minimizes frame error rate (FER) for a well-designed LDPC code.  However, this section also provides an example of a poorly-designed LDPC code where maximizing mutual information does not minimize FER.  In this example, larger erasure regions than would maximize the MI are needed to mitigate small absorbing sets.  The section concludes by presenting simulation results for these two LDPC codes using the channel model of \cite{DongUSENIX2011}. Section \ref{conclusion} delivers the conclusions.

The new material in this paper relative to the precursor conference paper \cite{JWANGGLOBECOM11} includes expressions for the derivatives of MI for two and three reads, application of the constant-ratio method to simplify optimization, comparison of the Gaussian model of unconstrained MLC optimization with optimization constrained to a single parameter $q$, use of a more realistic MLC channel model,  density-evolution and simulation results illustrating the trade-off of code performance across a range of quantization precisions, and both a demonstration of how maximizing MI can minimize FER and an example in which the performance of a poorly-designed code is {\em not} optimized by the quantization that maximizes the MI.


\begin{figure}
\centering
\includegraphics[width=0.4\textwidth]{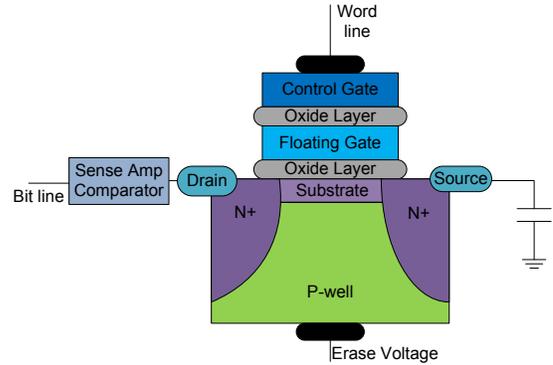}
\caption{A NAND Flash memory cell.}\label{flashcell}
\end{figure}

\section{The Read Channel of NAND Flash Memory}\label{background}

This paper focuses on the NAND architecture for Flash memory\deleted{, which is the most prevalent architecture today}. Fig. \ref{flashcell} shows the configuration of a NAND Flash memory cell. Each memory cell in the NAND architecture features a transistor with a control gate and a floating gate. To store information, a charge level is written to the cell by adding a specified amount of charge to the floating gate through Fowler-Nordheim tunneling by applying a relatively large voltage to the control gate \cite{BezIEEE03}.

To read a memory cell, the charge level written to the floating gate is detected by applying a specified word-line voltage to the control gate and measuring the transistor drain current. When reading a page, a constant word-line voltage is applied to all cells in the page.  A sense amp comparator compares the drain current  to a threshold. If the drain current is above this threshold, then the word-line voltage was sufficient to turn on the transistor.  This indicates that the charge written to the floating gate is below a certain value.  The sense amp comparator provides only one bit of information about the charge level present in the floating gate. 

A bit error occurring at this threshold-comparison stage is a \emph{raw bit error}  and the phrase \emph{channel bit error probability} refers to the probability of a raw bit error given a specified amount of distortion in the process of writing to the cell, retaining the charge level over a period of time, and reading the cell.  We refer to this overall process as the \emph{read channel}.

The word-line voltage or reference voltage required to turn on a particular transistor (called the threshold voltage) can vary from cell to cell for a variety of reasons. For example{, the floating gate can be overcharged during the write operation, the floating gate can lose charge due to leakage in the retention period, or the floating gate can receive extra charge when nearby cells are written \cite{MaedaISDFT09}.}  The variation of threshold voltage from its intended value is the \emph{read channel noise}.

We \added{initially} assume an i.i.d. Gaussian threshold voltage for each level of an SLC (i.e., two-level) Flash memory cell.  This is equivalent to binary phase-shift keying (BPSK) with additive white Gaussian noise (AWGN), except that the threshold voltage cannot be directly observed.  Rather, \added{at most} one bit of information about the threshold voltage may be obtained by each cell read.

More precise models such as the model in \cite{MaedaISDFT09}, in which the lowest and highest threshold voltage distributions have a higher variance, and the model in \cite{LiVLSI10}, in which the lowest threshold voltage (the one associated with zero charge level) is Gaussian and the other threshold voltages have Gaussian tails but a uniform central region, are sometimes used. The model in  \cite{DongUSENIX2011} is similar to \cite{LiVLSI10}, but is derived by explicitly accounting for {real dominating noise sources, such as inter-cell interference, program injection statistics, random telegraph noise and retention noise}.  \added{After considering  the simple Gaussian approximation for SLC, this paper considers MLC (four-level) Flash memory cells and uses the model of \cite{DongUSENIX2011} to study the maximum mutual information (MMI) approach and constant-ratio method in a more realistic setting to complement the analysis using a simple i.i.d. Gaussian assumption.}

In the next section, we {present a general quantization approach for selecting word-line voltages} for reading Flash memory cells and apply it to the specific example of SLC (two-level) Flash using a simple identically distributed Gaussian channel model.

\section{Soft Information Via Multiple Cell Reads} \label{casestudy}
Because the sense-amp comparator \deleted{only} provides \added{at most} one bit of information about the threshold voltage (or equivalently \added{about} the amount of charge present in the floating gate), decoders for error control codes in Flash have historically \replaced{used}{relied on} hard \deleted{bit} decisions \added{on each bit}.

\subsection{Obtaining Soft Information}
 \replaced{Soft}{However, soft} information can be obtained in two ways: either by reading from the same sense-amp comparator multiple times with different word-line voltages (as is already done to read multi-level Flash cells) or by equipping a Flash cell with multiple sense-amp comparators on the bit line, which is essentially equivalent to replacing the sense amp comparator (a one-bit A/D converter) with a higher-precision A/D converter.

These two approaches are not completely interchangeable in how they provide information about the threshold voltage. If the word-line voltage and floating gate charge level place the transistor in the linear gain region of the drain current vs. word-line-voltage curve (the classic I-V transistor curve), then valuable soft information is provided by multiple sense amp comparators.  However, multiple comparators may not give much additional information if the I-V curve is too nonlinear. If the drain current has saturated too low or too high, the outputs from more sense-amp comparators are not useful in establishing precisely how much charge is in the floating gate. 

In contrast, each additional read of a single sense amp comparator can provide additional useful information about the threshold voltage if the word-line voltages are well-chosen.  
Our work focuses on obtaining soft information from multiple reads using the same sense-amp comparator with different word-line voltages.  This approach was studied in \cite{DongTCAS2011}, and the poor performance of uniformly spaced word-line voltages was established.  

The fundamental approach of this paper is to choose the word-line voltages for each quantization by maximizing the MI between the input and output of the equivalent discrete-alphabet read channel.  This approach has been taken in other work (not in the context of Flash memory) such as \cite{LeeISIT05,KurkoskiIT}.   Theoretically, this choice of word-line voltages maximizes the amount of information provided by the quantization.  This section explores the simplest possible case, SLC (two-level) Flash using an identically distributed Gaussian model, which is equivalent to BPSK transmission with Gaussian noise.

\subsection{Quantizing Flash to Maximize Mutual Information}\label{sec:gaussian}

This subsection describes how to select word-line voltages to achieve maximum mutual information (MMI) for two reads and three reads for the identically distributed Gaussian model. 

\begin{figure}
\centering
\includegraphics[width=0.49\textwidth]{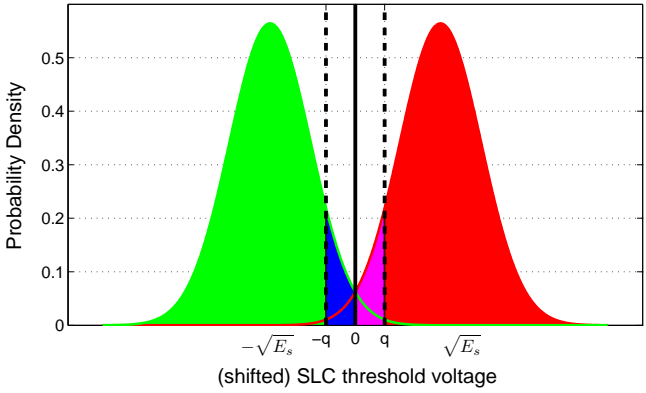}
\caption{Identically distributed Gaussian model for SLC threshold voltages.  Also shown are word-line voltages for two reads (the dashed lines) and three reads (all three lines).  The quantization regions are indicated by shading with the middle region for two reads being the union of the blue and purple regions.}\label{quantSLC2reads}\label{quantSLC3reads}
\end{figure}

For SLC Flash memory, each cell can store \replaced{one}{1} bit of information. Fig. \ref{quantSLC2reads} shows \replaced{the}{a simplistic} model of the threshold voltage distribution as a mixture of two  \added{identically distributed} Gaussian random variables.  When a  ``0''  or  ``1'' is written to the cell, the threshold voltage is modeled as a Gaussian random variable with variance $N_0/2$ and mean $-\sqrt{E_s}$ (for  ``1'' ) or mean $+\sqrt{E_s}$ (for  ``0'' ), respectively.

\subsubsection{Two reads per cell} 
\begin{figure}
\centering
\includegraphics[width=0.43\textwidth]{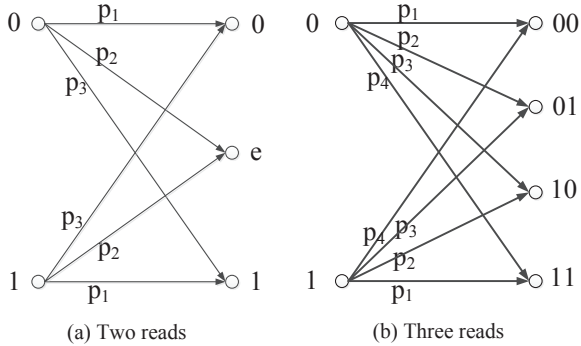}
\caption{Equivalent discrete memoryless channel models for SLC with (a) two reads and (b) three reads with distinct word-line voltages.}
\label{fig:slc2read3read}
\end{figure}

For SLC with two reads, Fig.~\ref{quantSLC2reads} shows symmetric word-line voltages $ q $ and $ -q $.  The threshold voltage is quantized into three regions as shown in Fig.~\ref{quantSLC2reads}: the green region,  the red region, and the union of the blue and purple regions (which essentially corresponds to an erasure $e$).  This quantization produces the effective discrete memoryless channel (DMC) model shown in Fig.~\ref{fig:slc2read3read}(a) with input  $ X \in \{0,1\}$ and output $ Y \in \{0,e,1\} $.


Assuming $ X $ is equally likely to be 0 or 1, the MI $I(X;Y)$ between  the input $X$ and output $Y$ of the resulting DMC can be calculated \cite{CoverEOIT} as
\begin{align}
I(X;Y) &= H(Y)-H(Y|X) \label{equ:MI2readsA}\\
&= H\left( \frac{\text{$p_{13}$}}{2},p_2,\frac{\text{$p_{13}$}}{2}\right)-H\left(p_1,p_2,p_3\right),\label{equ:MI2readsB}
\end{align}
where $H$ is the entropy function \cite{CoverEOIT}, $p_{ij}=p_i + p_j$,  and the crossover probabilities shown in Fig.~\ref{fig:slc2read3read}(a) are \mbox{$p_1=1-Q^{-}$, $p_2 = Q^{-} - Q^{+}$, and $p_3 =Q^{+}$}  with
\begin{equation}
Q^{-} = Q\left( \frac{\sqrt{E_s}-q}{\sqrt{N_0/2}} \right)\text{ and } Q^{+} = Q\left( \frac{\sqrt{E_s}+q}{\sqrt{N_0/2}}\right), \label{eq:Q-Q+}
\end{equation}
where $Q(x) = \frac{1}{\sqrt{2 \pi}} \int_x^{\infty}e^{-u^2/2}du$.

For fixed SNR $\frac{E_s}{N_0/2} $, the MI in (\ref{equ:MI2readsA}-\ref{equ:MI2readsB}) for the identically distributed Gaussian model is a quasi-concave function of $q$ with a zero derivative only at the $q$ that delivers the maximum MI and at $q=\infty$. The MI can be maximized analytically by setting $dI/dq =0$.  Let $f(x)$ be the probability density function of a standard normal distribution. The derivative is computed as
\begin{align}
\frac{dI}{dq} =f \left(T_q^+ \right)\log_2\left( \frac{p_{13}}{2p_3} \right) + f \left(T_{q}^- \right)\log_2 \left( \frac{p_{13}}{2p_1} \right) \label{eq:p1plusp3}, 
\end{align}
where $T_q^+ = \sqrt{E_s}+q$ and $T_{q}^- = \sqrt{E_s}-q$. 

Note that  $dI/dq$ is continuous on $\mathbb{R}^+$. At $q = 0$ we have $p_{13}= p_1+p_3 = 1$. Applying this to \eqref{eq:p1plusp3}, we have
\begin{align}
\frac{dI}{dq} = -f \left(\sqrt{E_s} \right)\log_2 \left( 4p_1(1-p_1) \right) \geq 0\,, \label{eq:dIdq-at-q=0}
\end{align}
at $q = 0$ by the inequality of arithmetic and geometric means.  Equality holds only when $p_1 = 1/2$, which also causes $I(X;Y)=0$.  It can also be shown that $dI/dq$  becomes negative for sufficiently large $q$  and then increases monotonically, approaching zero as $q$ approaches infinity.  

\begin{figure}
\centering
\includegraphics[width=0.52\textwidth]{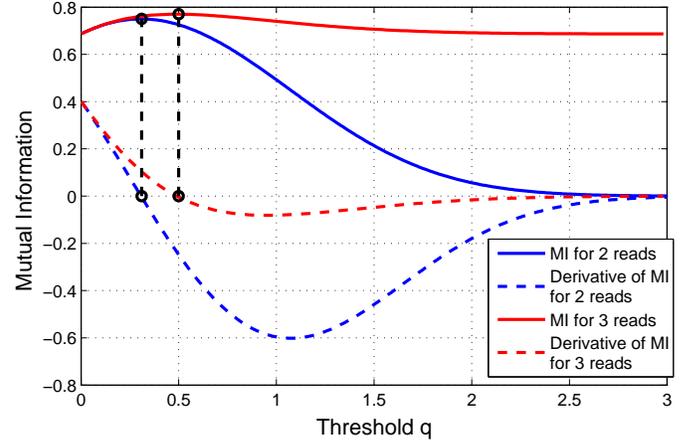}
\caption{MI and its derivative vs. $q$ (for $E_s=1$) for SNR$= \frac{E_s}{N_0/2}= 4$ dB for SLC (two-level) Flash with two reads and with three reads under the identically distributed Gaussian model.}\label{fig:MI_vs_q_2reads}
\end{figure}

These properties, illustrated in the example of Fig. \ref{fig:MI_vs_q_2reads}, ensure that there is a single zero derivative for finite $q$ corresponding to the desired maximum MI.  Because \eqref{eq:p1plusp3} involves the Q function, solving for the $q$ that sets $\frac{dI}{dq}=0$ requires a numerical approach such as the bisection search \cite{Covexoptimization}.

The blue curves in Fig. \ref{fig:MI_vs_q_2reads} show how the MI for two reads and its derivative vary as a function of $q$ for an SNR of 4 dB.   Note that when $q=0$ there is no erasure region, which is equivalent to a single read.  As $q$ increases so does the erasure region.  MI is concave in $q$ between $q=0$ and the point of inflection. Note that when $q=\infty$ the channel always produces the output $e$ and the MI is zero.


\subsubsection{Three reads per cell}
Now consider SLC with three reads for each cell.  The word-line voltages should again be symmetric (shown as $ q $, $0$, and $ -q $ in Fig.~\ref{quantSLC3reads}).  The threshold voltage is quantized according to the four differently shaded regions shown in Fig.~\ref{quantSLC3reads}.  This quantization produces the DMC model as shown in Fig.~\ref{fig:slc2read3read}(b) with input  $ X \in \{0,1\}$ and output $ Y \in \{00, 01, 10, 11\} $.  

Assuming $ X $ is equally likely to be 0 or 1, the MI between the input and output of this DMC can be calculated as
\begin{align}
I(X;Y) =& H(Y)-H(Y|X)\notag\\
=& H\left(\frac{p_{14}}{2},\frac{p_{23}}{2},\frac{p_{23}}{2},\frac{p_{14}}{2}\right) -H(p_1,p_2,p_3,p_4),\label{equ:mu3reads}
\end{align}
where $p_{ij}= p_i+p_j$ with $p_1 = 1-Q^{-}$, $p_2 = Q^{-} - Q^{0}$, $p_3 = Q^{0} - Q^{+}$, and $p_4 =Q^{+}$. $Q^{-}$ and $Q^{+}$ are as in \eqref{eq:Q-Q+} and 
\begin{equation}
 Q^{0} = Q\left( \frac{\sqrt{E_s}}{\sqrt{N_0/2}}\right).
\end{equation}

The derivative of MI with respect to the threshold q is 
\begin{align}
\dfrac{dI}{dq}=&\sum_{j = 1}^{4}p_{j}^{\prime}\log_2(p_j)-p_{14}^{\prime}\log_2(p_{14})-p_{23}^{\prime}\log_2(p_{23}) \, ,
\end{align}
where $-p_1^{\prime} = p_2^{\prime} = f(T_q^-)$  and $p_3^{\prime} = -p_4^{\prime} = f(T_q^+)$. When $q=0$, \eqref{eq:dIdq-at-q=0} still applies.

The red curves in Fig. \ref{fig:MI_vs_q_2reads} show how the MI for three reads and its derivative vary as a function of $q$ for an SNR of 4 dB. Both at $q=0$  and $q=\infty$ the channel is equivalent to the binary symmetric channel (BSC) produced by a single read with the threshold at zero.  Thus the MI for both of these extreme choices is identical.  Fig. \ref{fig:MI_vs_q_2reads} shows a single zero derivative corresponding to the desired maximum MI occurring between these two extremes.  Again, solving for the $q$ that sets $\frac{dI}{dq}=0$ requires a numerical approach such as the bisection search \cite{Covexoptimization}.

For the relatively simple identically distributed Gaussian model, ${dI}/{dq}$ for the two-read and three-read cases can be identified analytically as described above.  However, even in realistic models in which the distributions are described numerically, the optimum $q$ can usually be found by a bisection search.  Also, when the distributions are not identically distributed, the constant-ratio approach of \cite{DongTCAS2011}, which is introduced in Section \ref{sec:4-levels} for the four-level MLC case, can be used to produce a single-parameter optimization that again can be solved with quasi-convex optimization methods \cite{Covexoptimization}.

\subsection{Performance vs. Number of Reads Per Cell}
The MMI optimization approach generalizes to more than three reads per cell, but the optimization becomes more complex.  In these cases, there is more than one parameter and MI is not necessarily concave or quasi-concave in these parameters.  For these cases we used a coarse brute-force search of the parameter space  followed by a bisection optimization performed on promising small regions of the space until the optimal set of thresholds (within a small tolerance) was identified.  

\begin{figure}
\centering
\includegraphics[width=0.52\textwidth]{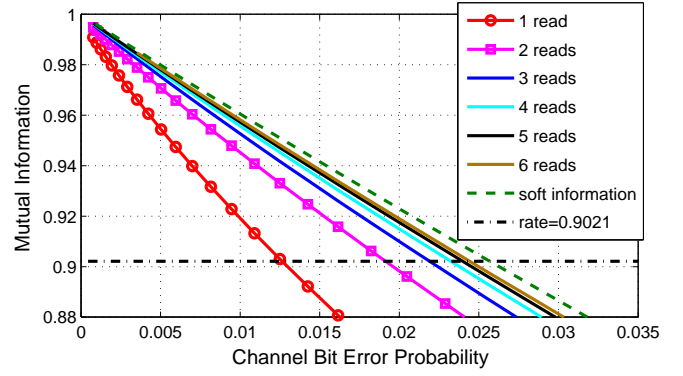}
\caption{MI provided by different quantizations for the identical Gaussian distributions model of SLC (two-level) Flash.  The dashed horizontal line indicates the operating rate of our simulations.  When an MI curve is below the dashed line, the read channel with that quantization cannot possibly support the attempted rate 0f 0.9021.}\label{mutual_slc}
\end{figure}

Fig.~\ref{mutual_slc} plots MI vs. channel bit error probability for  a range of number-of-reads-per-cell for the identical Gaussian distributions model of SLC.  MI increases with the number of reads.  The top (dashed) curve shows the MI possible with full soft information (where the decoder would know the threshold voltage exactly).  The bottom curve shows the MI available with a single read.   With two reads, the MI is improved enough to close about half of the gap between the single-read MI and the MI of full soft information.   Increasing the number of reads improves the MI, but with diminishing returns.  The bit error probability requirement to achieve an MI of 0.9021 (where the MI curve crosses the horizontal line in Fig.~\ref{mutual_slc}) increases (relaxes) as the number of reads increases.

\begin{figure}
\centering
\includegraphics[width=0.50\textwidth]{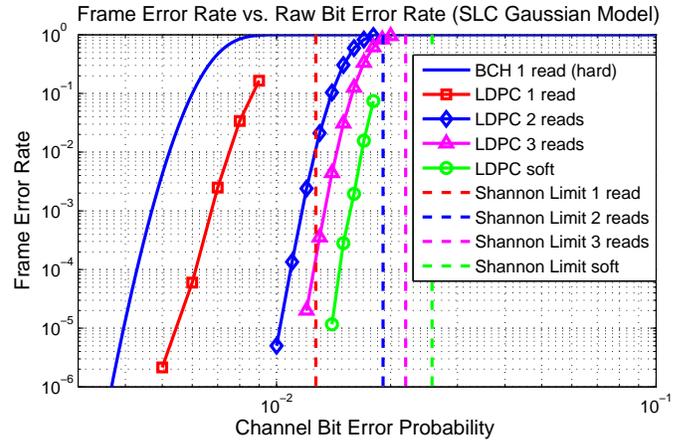}
\caption{Simulation results of FER vs. channel bit error probability using the Gaussian channel model for SLC (two-level) Flash comparing LDPC Code 2 with varying levels of soft information and a BCH code.  Both codes have rate 0.9021.  The BCH and LDPC 1-read curves correspond to hard decoding.}\label{slc_sim_code2}
\end{figure}

Fig. \ref{slc_sim_code2} shows how the performance of an LDPC code (Code 2 described in Section \ref{sec:LDPCDesign} below) improves as more soft information is made available to the decoder using MMI-optimized thresholds.  This simulation uses the Gaussian model of the  SLC Flash memory cell shown in Fig. \ref{quantSLC3reads}.  

Fig. \ref{slc_sim_code2} plots FER versus channel bit error probability computed as $Q \left( \sqrt {\frac{2 E_s}{N_0}} \right) $.  For reference, the FER performance of a binary BCH code capable of correcting up to 64 bit errors (using one read per cell) is also shown.  Both the LDPC code and the BCH code have rate 0.9021.   The LDPC code has a frame size of $k=8225$ bits and the BCH code has a frame size of $k=8256$ bits.  Also for reference, dashed vertical lines show the Shannon limit (worst channel that could theoretically support reliable transmission) for each level of quantization at the target rate of 0.9021.

Consistent with the mutual information curves of  Fig.~\ref{mutual_slc}, this plot illustrates that each additional read improves the FER performance of the LDPC code, but the performance improvement is diminishing.  Using three reads places the LDPC code performance within a relatively small gap from the limit of the performance achieved by that code with full soft information (essentially, an infinite number of reads).  Note that the LDPC code outperforms the BCH code even with a single read, but one or two additional reads significantly improve performance.

\section{LDPC Code Descriptions}\label{sec:LDPCDesign}

LDPC codes \cite{Gallagerthesis} are linear block codes defined by sparse parity-check matrices. By optimizing the degree distribution, it is well-known that LDPC codes can approach the capacity of an AWGN channel \cite{RichardsonDes}. Several algorithms have been proposed to generate LDPC codes for a given degree distribution, such as the ACE algorithm \cite{TianTCOM04} and the PEG algorithm \cite{PEG01}.

In addition to their powerful error-correction capabilities, another appealing aspect of LDPC codes is the existence of low-complexity iterative algorithms used for decoding. These iterative decoding algorithms are called belief-propagation algorithms. Belief-propagation decoders commonly use soft reliability information about the received bits, which can greatly improve performance. Conversely, a quantization of the received information which is too coarse can degrade the performance of an LDPC code.

Traditional algebraic codes, such as BCH codes, commonly use bounded-distance decoding and can \deleted{only} correct \added{up to} a specified, fixed number of errors. Unlike these traditional codes,  {it can be difficult for LDPC codes to } guarantee a specified number of correctable errors.  However the average bit-error-rate performance can often outperform that of BCH codes in Gaussian noise.

\subsection {Description of LDPC Codes}
In this paper we consider three irregular LDPC codes, which we will refer to as Code 1, Code 2, and Code 3.  These codes were selected to illustrate two points about LDPC codes in the context of limited-precision quantization.  The first point, illustrated later in this section, is that the relative performance of LDPC codes (i.e., which one is better) can depend on the level of quantization.   Codes 2 and 3 were selected so that Code 2 has a better density evolution threshold than Code 3 for a single read while Code 3 has a better density evolution threshold than Code 2 for the full-precision AWGN channel. 

The second point is that MMI quantization does not provide the right threshold for every code but should provide the right threshold as long as the code is good enough.  This point is explored in Section \ref{sec:4-levels}.  Code 1 provides an example of a code that is bad enough (because of small absorbing sets) that MMI quantization does not provide the correct quantization thresholds.  Code 2 is a well-designed code that avoids these absorbing sets and for which the MMI quantization minimizes the frame error rate.  

Codes 1 and 3 have degree distributions that optimize the density evolution threshold for the full-precision AWGN channel with maximum variable node degrees of 19 and 24 respectively. The degree distribution for Code 2 is a modification of the Code 1 degree distribution.  It was not explicitly designed to optimize any density evolution threshold, but has a better density evolution threshold for the single-read AWGN channel than either Code 1 or Code 3.

The LDPC matrices\footnote{The complete LDPC code parity-check matrices are available at the CSL website \url{www.seas.ucla.edu/csl/files/publications.html#COD}.}  were constructed according to their respective degree distributions using the ACE algorithm \cite{TianTCOM04} and the stopping-set check algorithm \cite{RamamoorthyICC04}.   All of the simulations were performed using a maximum of 50 iterations of a sequential belief propagation decoder.   Decoding stops as soon as all check nodes are satisfied. The frame size is $k=8225$ bits for each of the three LDPC codes.
The degree distributions of the three codes are as follows:
\begin{align*}
\lambda_1(x) = & 2.0054\times 10^{-5} + 3.5776 \times 10^{-2} x+ 0.39869x^2 \\
&+8.4827\times10^{-3}x^8 + 3.7701\times 10 ^{-2}x^9+0.51933x^{18}\\
\rho_1(x) =& 0.15662x^{54} + 0.84338x^{55}\\
\lambda_2(x) = & 1.7701\times 10^{-5} + 3.1579\times 10^{-2} x+ 0.46923x^3 \\
&+ 7.4877\times10^{-3}x^8 + 3.3278\times 10 ^{-2} x^9 + 0.45841x^{18}\\
\rho_2(x)=&1.0975\times 10 ^{-3}x^{61} + 0.73267x^{62} + 0.26623x^{63}\\
\lambda_3(x) = & 3.2172\times 10^{-2}x + 2.681\times10^{-3}x^2\\
&+ 0.55764x^3 + 0.40751x^{23}\\
\rho_3(x)=&0.10366x^{57} + 0.89634x^{58} \, ,\\
\end{align*}
where $\lambda(x)$ is the left (variable-node) degree distribution and $\rho(x)$ is the right (check-node) degree distribution.  A term of $a x^{d-1}$ in $\lambda(x)$ indicates that $a$ is the fraction of edges connecting to variable nodes with degree $d$.

\subsection{Quantization-based Design Trade-off}
Because reading a page of bits from the sense-amp comparators is a time-intensive operation, it is likely that enhanced precision will be added progressively only if needed to facilitate successful decoding.  Hence, a single LDPC code will be decoded at a variety of precision levels, which introduces a design trade-off that can be illustrated with two LDPC codes.

Table \ref{tbl:DensityEvolution} shows the density evolution thresholds for these three codes for the extremes of a full-precision SLC channel and a single-read SLC channel assuming the Gaussian model of Fig. \ref{quantSLC3reads}. Table \ref{tbl:DensityEvolution} reveals a trade-off between full precision performance and single-read performance.  For example, Code 2 has a lower (in dB) single-read threshold than Code 3, but a higher full-precision threshold than Code 3.   The density evolution differences indicate that the different channels produced by the different quantizations will typically have different optimal LDPC codes under iterative belief propagation decoding.

\begin{figure}
\centering
\includegraphics[width=0.49\textwidth]{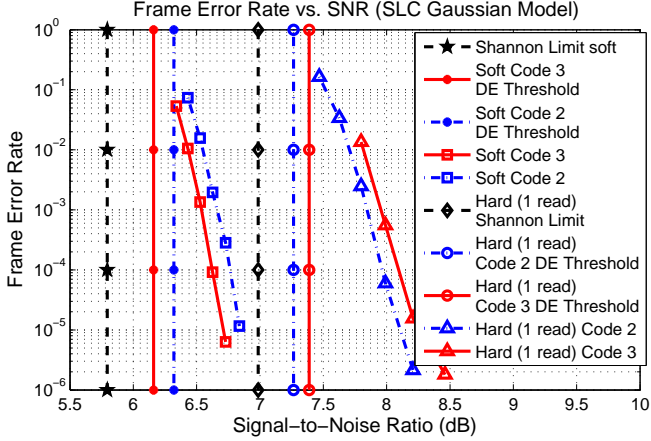}
\caption{Simulation results of FER vs. SNR$=2E_s/N_0$ using the Gaussian channel model for SLC (two-level) Flash comparing LDPC Code 2  and LDPC Code 3 with hard decoding (1 read) and full soft decoding (essentially an infinite number of reads).  Also shown are the Shannon limits for hard and soft decoding and the density evolution thresholds for the two codes under the two quantization scenarios.  Both the density evolution results and simulation results show a trade-off between performance under hard decoding and performance under soft decoding. }\label{PlotBPSKCode2_3}
\end{figure}

Fig. \ref{PlotBPSKCode2_3} shows FER vs. SNR simulation results consistent with the density evolution threshold results shown in Table \ref{tbl:DensityEvolution}.  Code 3 outperforms Code 2 when full soft information is available, but Code 2 outperforms Code 3 when only a single read is available.  Additional FER vs. channel BER simulations for Codes 2 and 3 (that were omitted from Fig. \ref{PlotBPSKCode2_3} for clarity of presentation) demonstrate that for 2 reads the codes have essentially the same performance, but for three reads Code 3 has better performance than Code 2.  
\begin{table}
\caption{Density evolution thresholds for three LDPC codes.  Full-precision threshold are in terms of both noise variance $\sigma$ and SNR.    Single-read thresholds are in terms of channel bit error probability $\epsilon$ and the corresponding SNR. \label{tbl:DensityEvolution}}
\begin{center}
\begin{tabular}{|l|l|l|l|l|l|}
\hline
&\multicolumn{2}{c|}{Full-precision AWGN}&\multicolumn{2}{c|}{Single-Read AWGN}\\
\hline
Code&$\sigma$&$SNR=2E_s/N_0$&$\epsilon$&$SNR(\epsilon)$\\
\hline \hline
1&0.499&6.04 dB&$9.29 \times 10^{-3}$& 7.44 dB\\
\hline
2&0.483&6.32 dB&$1.05 \times 10^{-2}$&7.26 dB\\
\hline
3&0.492&6.16 dB&$9.61 \times 10^{-3}$&7.39 dB\\

\hline
\end{tabular}
\end{center}
\end{table}

\section{Quantization for MLC (4-levels)} \label{sec:4-levels}
In this section, we extend the quantization approach to handle more than two levels, introduce a more realistic channel model, and present a method to reduce optimization complexity when there are more than two levels.

\subsection{MMI Quantization for MLC} 
\begin{figure}[t]
\centering
\includegraphics[width=0.48\textwidth]{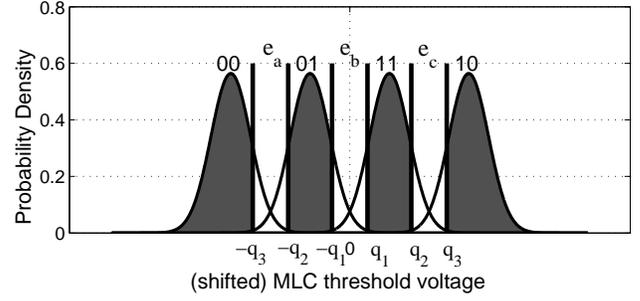}
\caption{Channel model for four-level MLC with threshold voltages modeled as Gaussians with the same variance.  Quantization is shown for six reads.}\label{quant4MLC6reads}
\end{figure}

For MLC (4-level) Flash memory, each cell can store 2 bits of information. Figure~\ref{quant4MLC6reads} extends the previously introduced SLC Gaussian model in the natural way.  Gray labeling $ (00,01,11,10) $ minimizes the raw bit error rate for these {four} levels. Typically in 4-level MLC Flash, each cell is compared to 3 word-line voltages and thus the output of the comparator has 4 possible values (i.e., four distinct quantization regions). 

If we consider three additional word-line voltages (for a total of six), the threshold voltage can be quantized to seven distinct regions as shown in Figure~\ref{quant4MLC6reads}.  The resulting DMC with four inputs and seven outputs is the natural extension of the DMCs shown in Fig.~\ref{fig:slc2read3read}.   In order to choose the optimal quantization levels $ q_1,q_2$, and $q_3 $ for a fixed SNR, we maximize the MI  which is computed as in \eqref{equ:MI2readsB} and \eqref{equ:mu3reads}, but with more terms.

The two bits corresponding to a single MLC cell are actually associated with two distinct pages in many Flash implementations. With Gray labeling as in Fig. \ref{quant4MLC6reads}, the most significant bit can be ascertained with a single read (or the two central reads for enhanced precision as shown in Fig. \ref{quant4MLC6reads}) without performing the other reads.  Similarly, the least significant bit using the labeling of Fig. \ref{quant4MLC6reads} can be ascertained from the two outer edge reads (or four outer edge reads for enhanced precision as shown in Fig. \ref{quant4MLC6reads})  without performing the central read(s).  

Because the read(s) associated with one of the two distinct bits turn out to be independent of the value of the other bit, the quantization optimization is not affected by whether the bits are stored in separate pages or not.   However, it should be noted that with Gray labeling as in Fig. \ref{quant4MLC6reads} the most significant bit enjoys a lower BER than the least significant bit for a given SNR.  In our LDPC simulations, a single binary LDPC code included both the most significant bit and the least significant bit of the relevant cells.  The inputs to the decoder are the reliabilities of these bits.

%

\begin{figure}[t]
\centering
\includegraphics[width=0.5\textwidth]{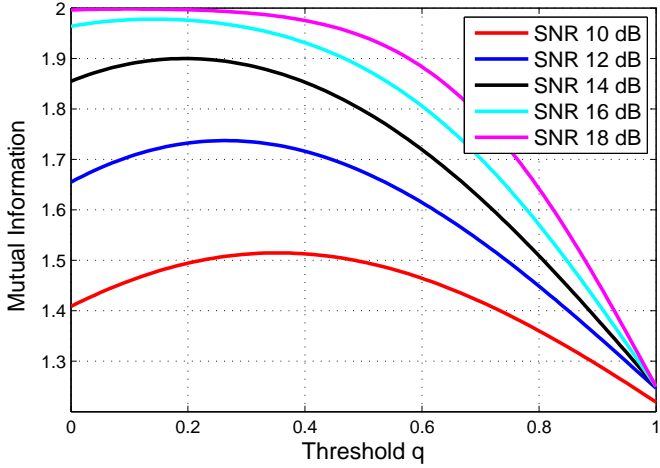}
\caption{MI vs. $q$ for various SNRs for the Gaussian model of MLC (four-level) Flash with the erasure regions in Fig. \ref{quant4MLC6reads} of size $2q$ and centered on the natural hard-decoding thresholds for Gaussians with means $\{\mu_1,\mu_2,\mu_3,\mu_4\} = \{-3,-1,1,3\}$.}\label{fig:FourLevelSingleq}
\end{figure}

The quantization problem can be constrained to a single parameter $q$ by selecting thresholds so that the three erasure regions in Fig. \ref{quant4MLC6reads} have the same size $2q$ and are centered on the natural hard-decoding thresholds.  With this constraint the problem becomes quasi-concave (or even concave) over the interesting region of $0 \le q \le (\mu_i - \mu_{i-1})/2$ as in Fig. \ref{fig:FourLevelSingleq}.

\begin{figure}[t]
\centering
\includegraphics[width=0.495\textwidth]{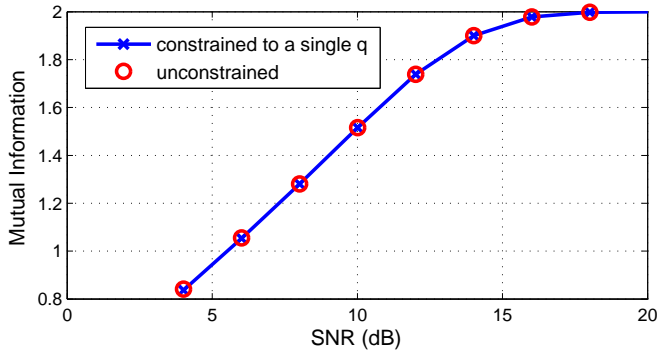}
\caption{MI vs. SNR for thresholds for MLC with six reads where optimization is either constrained by a single parameter $q$ or fully unconstrained.}\label{fig:ComparesingleqMMI}
\end{figure}

As we will see in Section \ref{sec:constant-ratio}, small differences in mutual information can lead to significant variations in FER.  Thus, it is important to understand whether the constrained thresholds studied in Fig. \ref{fig:FourLevelSingleq} cause a significant reduction in MMI as compared to unconstrained thresholds.  Fig. \ref{fig:ComparesingleqMMI} compares the performance of the constrained optimization, which has a single parameter $q$, and the full unconstrained optimization.  As shown in the figure, the benefit of fully unconstrained optimization is insignificant. 

\begin{figure}
\centering
\includegraphics[width=0.51\textwidth]{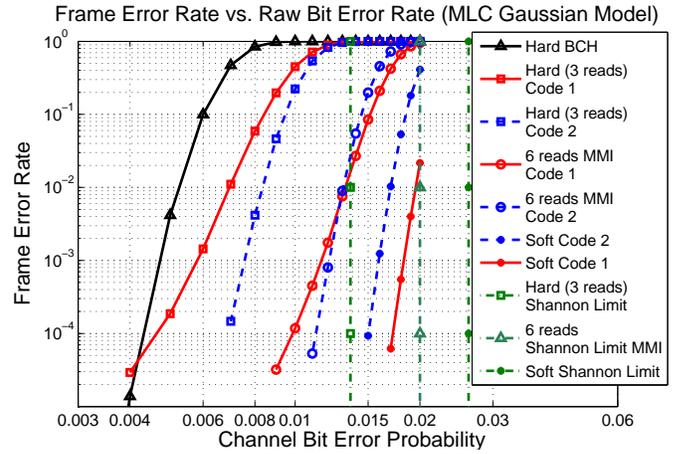}
\caption{FER vs. channel bit error probability simulation results using the Gaussian channel model for 4-level MLC comparing LDPC Codes 1 and 2 with varying levels of soft information and a BCH code with hard decoding.  All codes have rate 0.9021. }\label{sim:mlc4}
\end{figure}

Fig. \ref{sim:mlc4} shows performance of unconstrained MMI quantization on the Gaussian channel model of Fig. \ref{quant4MLC6reads} for three and six reads for Codes 1 and 2.  With four levels, three reads are required for hard decoding.  For MLC (four-level) Flash, using six reads recovers more than half of the gap between hard decoding (three reads) and full soft-precision decoding.  This is similar to the performance improvement seen for SLC (two-level) Flash when increasing from one read to two reads.

Note that in Fig. \ref{sim:mlc4}, the trade-off between performance with soft decoding and performance with hard decoding is even more pronounced.  Code 1 is clearly superior with soft decoding but demonstrates a noticeable error floor when decoded with three or six reads.  

\begin{figure}[t]
\centering
\includegraphics[width=0.3\textwidth]{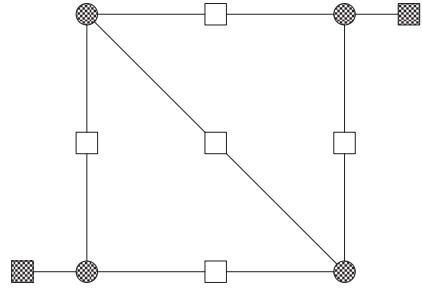}
\caption{A (4,2) absorbing set.  Variable nodes are shown as black circles.  Satisfied check nodes are shown as white squares.   Unsatisfied check nodes are shown as black squares.  Note that each of the four variable nodes has degree three.  This absorbing set is avoided by precluding degree-3 nodes. }\label{fig:abs42}
\end{figure}

LDPC error floors due to absorbing sets can be sensitive to the quantization precision, occurring at low precision but not at high precision \cite{dolecekIT10,WangIT}.   Code 1 has small absorbing sets including the $(4,2)$, $(5,1)$, and $(5,2) $ absorbing sets.  As shown in Fig. \ref{fig:abs42} for the (4,2) absorbing set, these absorbing sets can all be avoided by precluding degree-three variable nodes.  Code 2 avoids these absorbing sets because it has no degree-3 variable nodes.  As shown in Fig. \ref{sim:mlc4}, Code 2 avoids the error floor problems of Code 1.

\begin{figure}
\centering
\includegraphics[width=0.49\textwidth]{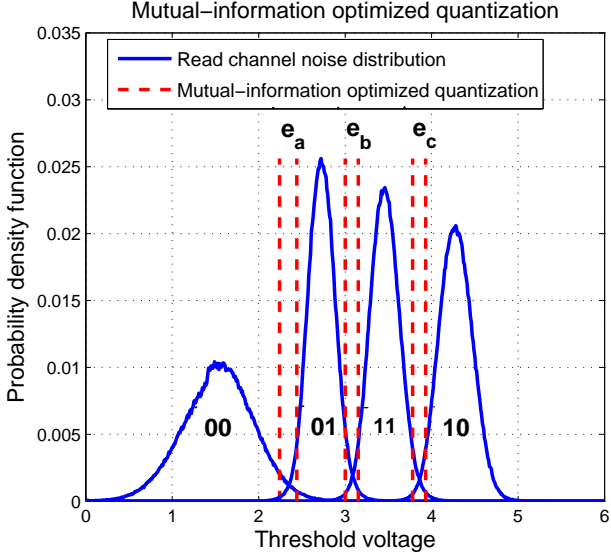}
\caption{Read channel noise distribution and mutual-information optimized quantization for six reads using the six-month retention model of \cite{DongUSENIX2011}.}\label{fig:pdf}
\end{figure}

\subsection{A More Realistic Model}\label{sec:retention}
We can extend the MMI analysis of Section~\ref{sec:gaussian} to any model for the Flash memory read channel. Consider again the 4-level 6-read MLC as a 4-input 7-output DMC.  Instead of assuming Gaussian noise distributions as shown in Fig. \ref{quant4MLC6reads}, Fig.~\ref{fig:pdf} shows the {four conditional threshold-voltage} probability density functions generated {according to}  the six-month retention model of \cite{DongUSENIX2011} and the {six} word-line voltages that maximize MI for this noise model.  {While the conditional noise for each transmitted (or written) threshold voltage is similar to that of a Gaussian, the variance of the conditional distributions varies greatly across the four possible threshold voltages.  Note that the lowest threshold voltage has by far the largest variance.}

Since the retention noise model is not symmetric, we need to numerically compute the transition probabilities and calculate the MI between the input and output as in  \eqref{equ:mu6reads_asym}.  

\begin{figure}[h]
\begin{align}
I(X;Y)=& H(Y)-H(Y|X)\notag\\
=& H\left(\frac{p_{11}+p_{21}+p_{31}+p_{41}}{4},\frac{p_{12}+p_{22}+p_{32}+p_{42}}{4}, \right. \notag\\
&\frac{p_{13}+p_{23}+p_{33}+p_{43}}{4},\frac{p_{14}+p_{24}+p_{34}+p_{44}}{4},\notag\\
&\frac{e_{1a}+e_{2a}+e_{3a}+e_{4a}}{4},\frac{e_{1b}+e_{2b}+e_{3b}+e_{4b}}{4},\notag\\
&\left. \frac{e_{1c}+e_{2c}+e_{3c}+e_{4c}}{4}\right)\notag\\
&-\frac{1}{4}H(p_{11},p_{12},p_{13},p_{14},e_{1a},e_{1b},e_{1c})\notag\\
&-\frac{1}{4}H(p_{21},p_{22},p_{23},p_{24},e_{2a},e_{2b},e_{2c})\notag\\
&-\frac{1}{4}H(p_{31},p_{32},p_{33},p_{34},e_{3a},e_{3b},e_{3c})\notag\\
&-\frac{1}{4}H(p_{41},p_{42},p_{43},p_{44},e_{4a},e_{4b},e_{4c}).\label{equ:mu6reads_asym}
\end{align}
\end{figure}

The MI in \eqref{equ:mu6reads_asym} is in general not a quasi-concave function in terms of the word-line voltages $ q_1,q_2,...,q_6 $.  Since \eqref{equ:mu6reads_asym} is a continuous and smooth function and locally quasi-concave in the range of our interest, we can numerically compute the MMI quantization levels with a careful use of bisection search or other quasi-convex optimization techniques \cite{Covexoptimization}.

\subsection{The Constant-Ratio Method}\label{sec:constant-ratio}

In \cite{DongTCAS2011},  a helpful heuristic constrains the additional word-line voltages to the left and right of each hard-decision word-line voltage so that the largest and second-largest conditional noise pdfs have a specified constant ratio $R$.  This is a natural extension to general non-symmetric channels of the constraint  to a single parameter by selecting thresholds so that the three erasure regions have the same size $2q$ and are centered on the natural hard-decoding thresholds in the simple symmetric Gaussian model of Fig. \ref{quant4MLC6reads}.

Note that the value of $R$ at the natural hard-decision threshold is one because the two densities are equal.  Higher values of $R$ move these secondary thresholds further away from the hard decoding thresholds.  In Fig. \ref{quant4MLC6reads} a higher value of $R$ would correspond to larger ``erasure'' regions (shown in white).   Although this heuristic is not named in \cite{DongTCAS2011}, we will refer it as the ``constant-ratio'' (CR) method.   

In \cite{DongTCAS2011}, the specification of $R$ is left to empirical simulation.  By choosing $R$ to maximize MI, the CR method can be viewed as a constraint that can be applied to MMI optimization to reduce the search space.  
The CR method can also simplify optimization because, as shown for the single-$q$ constraint in Fig. \ref{fig:FourLevelSingleq}, MI is a quasi-concave function of $R$ in the region of interest for the MLC (four-level) symmetric Gaussian channel.

\begin{figure}
\centering
\includegraphics[width=0.48 \textwidth]{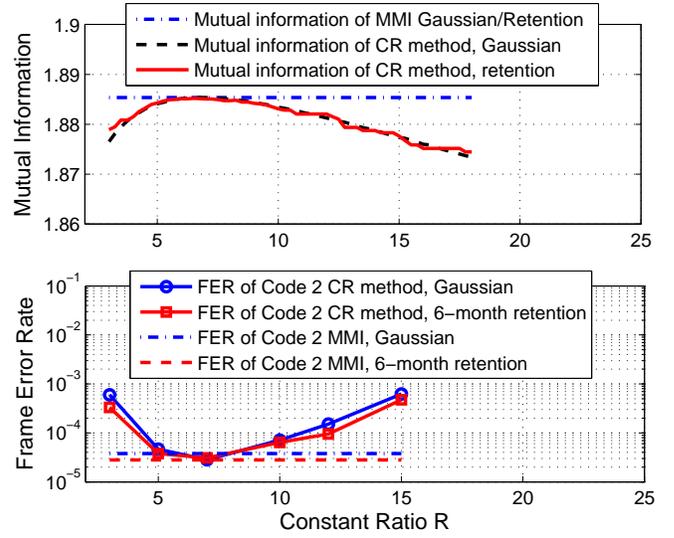}
\caption{Mutual information and frame error rate for Code 2 separately plotted as functions of the constant-ratio value $R$ for six quantization thresholds (seven levels).  Curves are shown for both the 4-PAM Gaussian model with SNR~$= 13.76$ dB and the retention model of \cite{DongUSENIX2011} for 6 months.  These two models both have an MMI of 1.885 bits, shown as a dashed line in the mutual information plot.  The frame error rate plots are for LDPC Code 2 described in Section \ref{sec:LDPCDesign}.  The two models had slightly different frame error rates with MMI quantization, $3.78 \times 10^{-5}$ for the 4-PAM Gaussian model and $2.8 \times 10^{-5}$ for the retention model, shown as dashed lines in the frame error rate plot.}\label{fig:MIvsR}
\end{figure}

Fig.~\ref{fig:MIvsR} shows MI as a function of $R$ for MLC (four-level) Flash with six quantization thresholds (seven quantization levels) for both the simple symmetric Gaussian model and the more realistic retention model of \cite{DongUSENIX2011}.   The Gaussian and retention channels were selected so that they have an identical MMI for six-read (seven-level) unconstrained MMI optimization.  

For both models the CR method with the MI-maximizing $R$ provides essentially the same MI as obtained by the unconstrained MMI optimization.  Furthermore, it is striking how similar the MMI vs. $R$ behavior is for the two different channel models.   For the Gaussian model, MI is a concave function of $R$.  The curve of MI vs. $R$ for the retention model closely follows the Gaussian model curve, but is not a strictly concave function because of variations in the numerical model.  


The MMI approach is a way to select quantization levels in the hope of optimizing frame-error-rate (FER) performance.  Fig.~\ref{fig:MIvsR} shows the FER performance as a function of $R$ for both the Gaussian model and the retention model for LDPC Code 2 described in Section \ref{sec:LDPCDesign}.  The value of $R$ that provides the maximum MI also delivers the lowest FER as a function of $R$.  This lends support to the approach of selecting quantization thresholds to maximize MMI.  

The constraint to a constant ratio does not appear to adversely affect FER; the lowest FER as a function of $R$ is essentially the same as the FER achieved by unconstrained MMI quantization.  The range of MI in Fig.~\ref{fig:MIvsR} is small (approximately 0.01 bits), but this variation in MI corresponds to more than an order of magnitude of difference in FER. 

\begin{figure}
\centering
\includegraphics[width=0.49 \textwidth]{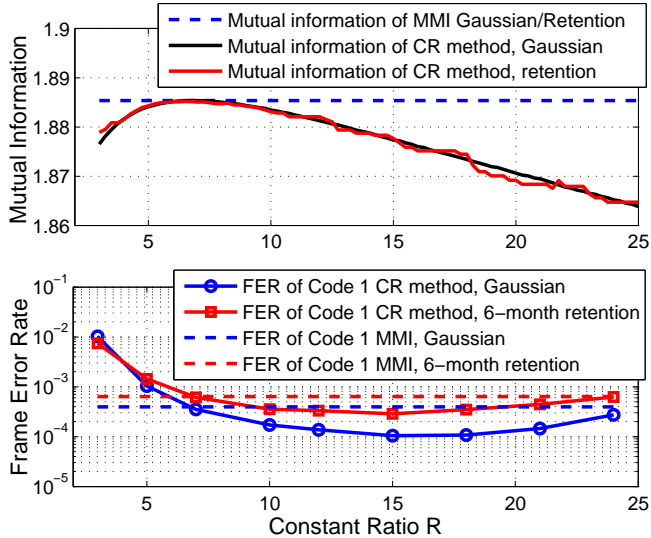}
	\caption{Mutual information and frame error rate for Code 1 separately plotted as functions of the constant-ratio value $R$ for six quantization thresholds (seven levels).  Curves are shown for both the 4-PAM Gaussian model with SNR~$= 13.76$ dB and the retention model of \cite{DongUSENIX2011} for 6 months.  These two models both have an MMI of 1.885 bits shown as a dashed line in the mutual information plot.  The frame error rate plots are for LDPC Code 1 described in Section \ref{sec:LDPCDesign}.}  \label{fig:MIvsR_code1}
	\vspace{-0.1in}
\end{figure}

\subsection{MMI Optimization Thwarted by Small Absorbing  Sets} \label{sec:MMI_opt}

While the previous example showed that maximizing MI can also minimize FER for a well-designed code, it is important to note that poorly designed codes can perform best with a quantization that does not maximize the channel MI.  

To illustrate this, we previously introduced Code 1, which has a high error floor under hard decoding due to the presence of numerous small absorbing sets.  As shown in Fig. \ref{fig:MIvsR_code1}, for Code 1, the lowest FER occurs with $R=15$ which provides less mutual information than $R=7$.   

This behavior may appear to be counter-intuitive.  However, the numerous small absorbing sets serve as traps that can turn a few hard-decoded errors into uncorrectable problems.  The presence of these absorbing sets forces the code to prefer a wider erasure region (thereby minimizing hard-decoded errors that trigger the absorbing sets) than would be optimal in terms of capacity.

\subsection{Simulation Results for Retention Model}

\begin{figure}[t]
\centering
\includegraphics[width=0.48\textwidth]{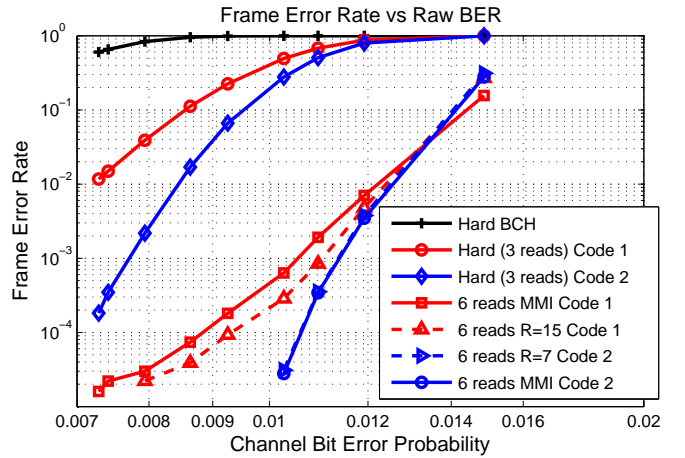}
\caption{FER vs. channel bit error probability results using the six-month retention model of \cite{DongUSENIX2011} for 4-level MLC.  All codes have rate 0.9021. Hard decoding results are shown for the BCH code and LDPC Codes 1 and 2.   FER performance for enhanced precision decoding using six reads is shown for LDPC Codes 1 and 2 using both unconstrained MMI quantization and MMI quantization with the constant-ratio constraint  with the $R$ value that minimizes FER for that LDPC code.  }\label{fig:simretention_code2}
\end{figure}

Now we examine code performance using the retention model of \cite{DongUSENIX2011}.  Fig.~\ref{fig:simretention_code2} shows frame error rate (FER) plotted versus retention time for Codes 1 and 2 with three reads and with six reads.  Code 2 outperforms Code 1 under both three reads (hard decoding) and six reads.

The  three-read quantization whose performance is shown in Fig. \ref{fig:simretention_code2} is  standard hard decoding for four-level MLC.   We note that in principle, since the retention model is not symmetric, some gain can be achieved by allowing asymmetric thresholds and optimizing these thresholds using MMI even in the three-read case.  However, we found those gains to be insignificant in our simulations.

In Fig. \ref{fig:simretention_code2}, the Code-2 FER curves for unconstrained-MMI quantization and for $R=7$ are indistinguishable.  Recall from Fig.~\ref{fig:MIvsR} that $R=7$ both maximizes the mutual information and minimizes frame error rate for Code 2.  This was the hoped-for result of MMI optimization, that it would also optimize the true objective of minimizing FER.  However, as we saw in Section \ref{sec:MMI_opt}, if an LDPC code has a high error floor, optimizing the MMI does not necessarily minimize the FER.  

Thus, a code with relatively poor performance can perform slightly better with a quantization that does not maximize the mutual information.  Indeed, the best FER performance for Code 1 in Fig. \ref{fig:MIvsR} for six reads with constant ratio quantization is with $R=15$.  Note from Fig.  \ref{fig:MIvsR_code1} that $R=15$ provides a smaller mutual information than $R=7$, but $R=15$ provides the lowest FER for Code 1.  

Notice in Fig. \ref{fig:simretention_code2} that for Code 1 with six reads, the MMI quantization performs slightly worse than the $R=15$ quantization.  Thus we can see that for a weaker code, the MMI approach may not provide the best possible quantization in terms of FER.  However, this situation may well be interpreted as an indicator that it may be worth exploring further code design to improve the code rather than adopting a different threshold optimization approach.

\section{Conclusion}\label{conclusion}
This paper explores the benefit of using soft information in an LDPC decoder for NAND Flash memory. Using a small amount of soft information improves the performance of LDPC codes significantly and demonstrates a clear performance advantage over conventional BCH codes. 

In order to maximize the performance benefit of the soft information, we present an approach for optimizing word-line-voltage selection so that the resulting quantization maximizes the mutual information between the input and output of the equivalent read channel. This method can be applied to any channel model.  Constraining the quantization using the constant-ratio method provides a significant simplification with no noticeable loss in performance.   Furthermore, only a few additional reads can harvest most of the performance improvement available through enhanced precision.

Our simulation results suggest that if the LDPC code is well designed, the quantization that maximizes the mutual information will also minimize the frame error rate.  However,  the MMI approach can fail to identify the lowest-FER quantization for an LDPC code with a high error floor.

Separately, an LDPC code degree distribution designed for a full-precision Gaussian channel may be sub-optimal in the quantized setting and vice versa.  A second and distinct design factor is that absorbing sets become more important as the precision of the soft information decreases.  Considering these two distinct effects, it would be useful to design a code that is optimal over a large range of precisions or to show that such universal performance is not possible.

In this paper, the channel information has been quantized with various levels of precision.  However, the messages passed between the variable nodes and check nodes of the decoder have been represented as floating point numbers in our simulations.  It is an interesting area of further investigation to consider limited-precision representations within the LDPC decoder in conjunction with the limited-precision channel information that is available in the Flash memory setting.




\bibliographystyle{unsrt}	

\bibliography{myrefs}		

\begin{IEEEbiographynophoto}{Jiadong Wang}
Jiadong Wang received his B.S. degree in automation from Tsinghua University, Beijing, P.R.China in 2007, and his M.S. and Ph.D. degrees in electrical engineering from University of California at Los Angeles, in 2008 and 2012, respectively. His research is in the area of communication theory with a focus on channel coding. He has worked on a broad range of research topics including LDPC codes and turbo codes for applications such as flash memory and broadcast channels. He is currently with Qualcomm Inc., San Diego, CA, where he conducts research on physical layer communications for CDMA technologies.
\end{IEEEbiographynophoto}

\begin{IEEEbiographynophoto}{Kasra Vakilinia}
Kasra Vakilinia is a Ph.D. student in Electrical Engineering Department at the University of California , Los Angeles (UCLA).  He is a member of the UCLA Communication Systems Laboratory (CSL) and UCLA Center on Development of Emerging Storage Systems (CoDESS). His research interests include coding theory, information theory, flash memory storage systems, and communications systems.
\end{IEEEbiographynophoto}

\begin{IEEEbiographynophoto}{Tsung-Yi Chen (S'11-M'13)}
Tsung-Yi Chen received his B.S. degree in Electrical Engineering from National Tsing Hua University, Taiwan, in 2007. He obtained his M.S. and Ph.D. degrees in Electrical Engineering from UCLA in 2009 and 2013, respectively. He is a recipient of the UCLA Dissertation Year Fellowship 2012-2013). He then joined Northwestern University, Evanston, in 2013 as a postdoctoral fellow. His research includes coding theory and information theory, with applications to feedback communication, flash memory storage systems, and machine-to-machine communication. 
\end{IEEEbiographynophoto}

\begin{IEEEbiographynophoto}{Thomas Courtade (S'06-M'13)}
Thomas Courtade received the B.S. degree in Electrical Engineering from Michigan Technological University in 2007, and the M.S. and Ph.D. degrees in Electrical Engineering from UCLA in 2008 and 2012, respectively.  In 2012, he was awarded the inaugural Postdoctoral Research Fellowship through the NSF Center for Science of Information. He is currently an Assistant Professor at the University of California at Berkeley. His honors include a Distinguished Ph.D. Dissertation award and an Excellence in Teaching award from the UCLA Department of Electrical Engineering and a Best Student Paper 
Award at the 2012 International Symposium on Information Theory.  
\end{IEEEbiographynophoto}

\begin{IEEEbiographynophoto}{Guiqiang Dong (S'09)}
Guiqiang Dong (S'09) received the B.S. and M.S. degrees from the University of Science and Technology of China, Hefei, China, in 2004 and 2008, respectively, and the Ph.D. degree from the Electrical, Computer and Systems Engineering Department, Rensselaer Polytechnic Institute, Troy, NY, USA, in 2012. He has discovered a novel method by successfully building a software tool to rapidly estimate the error floor of low-density parity-check codes. His research interests include coding theory, NAND Flash memory, error correction code and signal processing application for NAND Flash SSD, and firmware design for NAND Flash SSD. He is currently a Chief Channel Architect with Skyera, Inc., San Jose, CA, USA.
\end{IEEEbiographynophoto}

\begin{IEEEbiographynophoto}{Hari Shankar}
Hari Shankar received his BTech degree from IIT Kanpur in 1983 and his PhD from the University of Hawaii in 1990.  He is currently a principal engineer at Inphi Corporation.  He has worked on a variety of projects which include flash memory, optical communication systems and high speed data links.  Previously he has worked at Intel and Hewlett Packard on wireless communication systems and EDA software.  His interests are in high-speed communication system design.
\end{IEEEbiographynophoto}

\begin{IEEEbiographynophoto}{Tong Zhang (M'02-SM'08) }
Tong Zhang received the B.S. and M.S. degrees in electrical engineering from the Xian Jiaotong University, Xian, China, in 1995 and 1998, respectively. He received the Ph.D. degree in electrical engineering from the University of Minnesota, Minneapolis, in 2002. Currently he is a Professor in electrical, computer and systems engineering department at Rensselaer Polytechnic Institute, Troy, NY. His current research interests include circuits and systems for memory and data storage, computing, and signal processing. He has served as an Associate Editor for the IEEE Transactions on Circuits and Systems - II, and the IEEE Transactions on Signal Processing.
\end{IEEEbiographynophoto}

\begin{IEEEbiographynophoto}{Richard Wesel SM'91-M'96-SM'01 }
Richard D. Wesel is a Professor with the UCLA Electrical Engineering Department and
is the Associate Dean for Academic and Student Affairs for the UCLA Henry Samueli 
School of Engineering and Applied Science. He joined UCLA in 1996 after receiving 
his Ph.D. in electrical engineering from Stanford. His B.S. and M.S. degrees in electrical 
engineering are from MIT. His research is in the area of communication theory with 
particular interest in channel coding. He has received the National Science Foundation 
(NSF) CAREER Award, an Okawa Foundation award for research in information theory 
and telecommunications, and the Excellence in Teaching Award from the Henry Samueli 
School of Engineering and Applied Science.
\end{IEEEbiographynophoto}

\end{document}